\documentclass[sigconf,screen]{acmart}
\AtBeginDocument{%
  \providecommand\BibTeX{{%
    \normalfont B\kern-0.5em{\scshape i\kern-0.25em b}\kern-0.8em\TeX}}}

\newcommand{\rqone}{Do the chain of VQs repair the bugs in LLM-generated code?}
\newcommand{\rqtwo}{Can VQs introduce defects in LLM-generated code?}
\newcommand{\rqthree}{
How does rephrasing the template of VQs impact the performance of LLM in repairing the bugs?}

\usepackage{listings}
\definecolor{codegreen}{rgb}{0,0.6,0}
\definecolor{codegray}{rgb}{0.5,0.5,0.5}
\definecolor{codepurple}{rgb}{0.58,0,0.82}
\definecolor{backcolour}{rgb}{0.9,0.9,0.9}
\definecolor{red}{RGB}{160,0,0}

\lstdefinestyle{mystyle}{
    backgroundcolor=\color{backcolour},   
    commentstyle=\color{codegreen},
    keywordstyle=\color{magenta},
    numberstyle=\tiny\color{codegray},
    stringstyle=\color{codepurple},
    basicstyle=\ttfamily\footnotesize,
    breakatwhitespace=false,         
    breaklines=true,  
    captionpos=b,                    
    keepspaces=true,                 
    numbers=left,                    
    numbersep=5pt,                  
    showspaces=false,                
    showstringspaces=false,
    showtabs=false,                  
    tabsize=2,
}

\setcopyright{acmlicensed}
\acmDOI{10.1145/3664646.3664772}
\acmYear{2024}
\copyrightyear{2024}
\acmSubmissionID{fsews24aiwaremain-p72-p}
\acmISBN{979-8-4007-0685-1/24/07}
\acmConference[AIware '24]{Proceedings of the 1st ACM International Conference on AI-Powered Software}{July 15--16, 2024}{Porto de Galinhas, Brazil}
\acmBooktitle{Proceedings of the 1st ACM International Conference on AI-Powered Software (AIware '24), July 15--16, 2024, Porto de Galinhas, Brazil}
\received{2024-04-05}
\received[accepted]{2024-05-04}
\begin{document}

\title{Chain of Targeted Verification Questions to Improve the Reliability of Code Generated by LLMs }
\author{Sylvain Kouemo Ngassom}
\affiliation{%
  \institution{Polytechnique Montréal}
  \city{Montreal}
  \country{Canada}
}
\email{sylvain.kouemo-ngassom@polymtl.ca}

\author{Arghavan Moradi Dakhel}
\affiliation{%
  \institution{Polytechnique Montréal}
  \city{Montreal}
  \country{Canada}
}
\email{arghavan.moradi-dakhel@polymtl.ca}

\author{Florian Tambon}
\affiliation{%
  \institution{Polytechnique Montréal}
  \city{Montreal}
  \country{Canada}
}
\email{florian-2.tambon@polymtl.ca}

\author{Foutse Khomh}
\affiliation{%
  \institution{Polytechnique Montréal}
  \city{Montreal}
  \country{Canada}
}
\email{foutse.khomh@polymtl.ca}




\begin{abstract}
  LLM-based assistants, such as GitHub Copilot and ChatGPT, have the potential to generate code that fulfills a programming task described in a natural language description, referred to as a prompt. The widespread accessibility of these assistants enables users with diverse backgrounds to generate code and integrate it into software projects. However, studies show that code generated by LLMs is prone to bugs and may miss various corner cases in task specifications. Presenting such buggy code to users can impact their reliability and trust in LLM-based assistants. Moreover, significant efforts are required by the user to detect and repair any bug present in the code, especially if no test cases are available. 
  In this study, we propose a self-refinement method aimed at improving the reliability of code generated by LLMs by minimizing the number of bugs before execution, without human intervention, and in the absence of test cases. Our approach is based on targeted Verification Questions (VQs) to identify potential bugs within the initial code. These VQs target various nodes within the Abstract Syntax Tree (AST) of the initial code, which have the potential to trigger specific types of bug patterns commonly found in LLM-generated code. Finally, our method attempts to repair these potential bugs by re-prompting the LLM with the targeted VQs and the initial code. Our evaluation, based on programming tasks in the CoderEval dataset, demonstrates that our proposed method outperforms state-of-the-art methods by decreasing the number of targeted errors in the code between 21\% to 62\% and improving the number of executable code instances to 13\%. 
\end{abstract}





\maketitle

\section{Introduction}
The rapid advancements in Generative Artificial Intelligence (GenAI) offer different opportunities for its application across various domains, including Software Engineering (SE)~\cite{tufano2024unveiling}. Large Language Models (LLMs) like GPT4~\cite{achiam2023gpt}, Codex~\cite{chen2021evaluating}, and Llama-2~\cite{touvron2023llama} that can generate code based on a given natural language description, referred to as a prompt, have significantly advanced the automatic code generation process through their ability to synthesize code. 
The widespread accessibility of LLM-based assistant tools, such as ChatGPT~\cite{wu2023brief}, GitHub Copilot~\cite{GitHubCopilot}, and Copilot Chat~\cite{Copilotchat} allows individuals with different backgrounds to generate code and contribute to open-source projects~\cite{kazemitabaar2023novices,wu2023autogen}. However, there are still significant gaps between the code generated by LLM-based assistant tools and the code written by human software developers in terms of comprehension and quality~\cite{spiess2024quality,vaithilingam2022expectation,liu2024your}. 


The range of bugs in code generated by LLMs may vary from those easily detectable by an IDE linter to those that are obscure and challenging to identify~\cite{liu2023refining,tambon2024bugs}. Presenting such buggy code to end users by LLM-based assistant tools 
may increase the efforts on the user's side to detect and repair these bugs, in a way that some users prefer to rewrite the code from scratch rather than incorporating suggestions from LLM-based assistant tools~\cite{vaithilingam2022expectation}. 
Also, users may, at times, not be aware of what a correctly generated solution should resemble, and the absence of test cases further complicates evaluating LLM-generated code and detecting their bugs~\cite{liu2024your,agarwal2024copilot}. Thus, detecting and repairing those bugs in LLM-generated code before presenting them to the user can significantly improve their reliability. 


Crafting effective prompts can be a solution to reduce human involvement and improve the reliability of LLM-generated code~\cite{hassan2024rethinking}. However, available solutions, such as incorporating the error messages raised by the compiler into the prompt~\cite{schafer2023adaptive,skreta2023errors,liventsev2023fully} or using Automatic Programming Repair (APR) tools to repair the buggy code generated by LLM~\cite{fan2023automated,dakhel2023github}, require test cases to trigger bugs in the code fragment. But, test cases may not always be readily available. Some studies suggest that iterative refinement can enhance the quality of code generated by LLMs~\cite{jiang2023selfevolve,dakhel2023effective,skreta2023errors}. However, these refinement steps are generally applied and may not effectively address specific bugs in the generated code, or they may necessitate the presence of test cases~\cite{jiang2023selfevolve,liventsev2023fully}. 

In real software projects, developers typically pose verification questions to reduce ambiguity during the code review and improve the quality of submitted code~\cite{ebert2019confusion,han2021understanding}. Additionally, raising questions about unexpected or incorrect aspects of students' solutions helps them to think around the incorrect part of their solutions and fill the gaps~\cite{shaughnessy2021think}. 
Inspired by these studies, we introduce a self-refinement method that leverages LLMs to improve the reliability of LLM-generated code by asking a chain of targeted Verification Questions (VQ), in the absence of test cases or concrete execution of the code. 


Our method starts by converting the initial code generated by an LLM into an Abstract Syntax Tree (AST) \cite{cui2010code}. Then, the method extracts features from targeted nodes in the AST that have the potential to trigger specific bug patterns. We choose the targeted nodes based on their properties which make them likely to lead to particular bug patterns as presented in existing LLMs' generated code bug taxonomy \cite{tambon2024bugs}. Subsequently, our method incorporates features of targeted nodes into different VQ templates generated by an LLM, depending on the node types. The approach then makes use of ChatGPT (gpt3.5-turbo) to iteratively improve the reliability of the initial code by prompting the model with the chain of VQs and repairing potential bugs.  In the end, the method returns a repaired code where it fixed the targeted bug. The aim of our method, achieved through asking effective and targeted VQs, is to focus on a targeted bug while maintaining the correct code during the repairing process. In particular, our method does not require any prior execution of the code or a comprehensive set of test cases to repair the buggy code and improve the reliability of the initially generated code.

Our results indicate that our proposed method outperforms baseline methods, i.e., without any VQs or with general (non-targeted) VQs. In particular, our method reduces the amount of code generated with specific targeted error types up between 21\% to 62\% and improves the number of executable code instances generated by LLM by 13\%. Furthermore, our approach introduces relatively few new bugs, around 12\%, in the generated code (i.e. false positive) with the chain of VQs. Finally, while rephrasing the templates of targeted VQs can induce some variability in the output, the results remain consistent across rephrasing the template VQs differently in terms of their performance in bug repair.

To guide our study, we aim to answer the following RQs: 
\begin{itemize}
    \item \textbf{RQ1}: \rqone
    \item \textbf{RQ2}:  \rqtwo
    \item \textbf{RQ3:} \rqthree
\end{itemize}



The key contribution of this paper is:
\begin{itemize}
    \item Proposing a self-refinement method that generates a chain of targeted VQs to first localize and then repair specific bugs in LLM-generated code before execution, without human intervention, and in the absence of test cases. 
\end{itemize}

The targeted VQs in our study focus on the bug itself without replacing the entire code and it enables more control over the changes induced in the code during the repairing step. We are the first study that focuses on bug patterns that are prevalent in LLM-generated code. Our proposed method can be integrated as an agent that has been trained to enhance the reliability of LLM-generated code by generating and addressing a set of targeted VQs within an adversarial framework.

The rest of this paper is organized as follows: Section \ref{sec:meth} presents the methodology. Section \ref{sec:exp} describes the experimental setup along with the results. Section \ref{sec:discuss} discusses our results. Then, Section \ref{sec:ttv} details the threats to the validity of the study and Section \ref{sec:related} lists related works. Finally, the conclusion of the study is presented in Section \ref{sec:concl}.

\section{Methodology}\label{sec:meth}
\begin{figure*}[t]
\centerline{\includegraphics[width=0.9\textwidth]{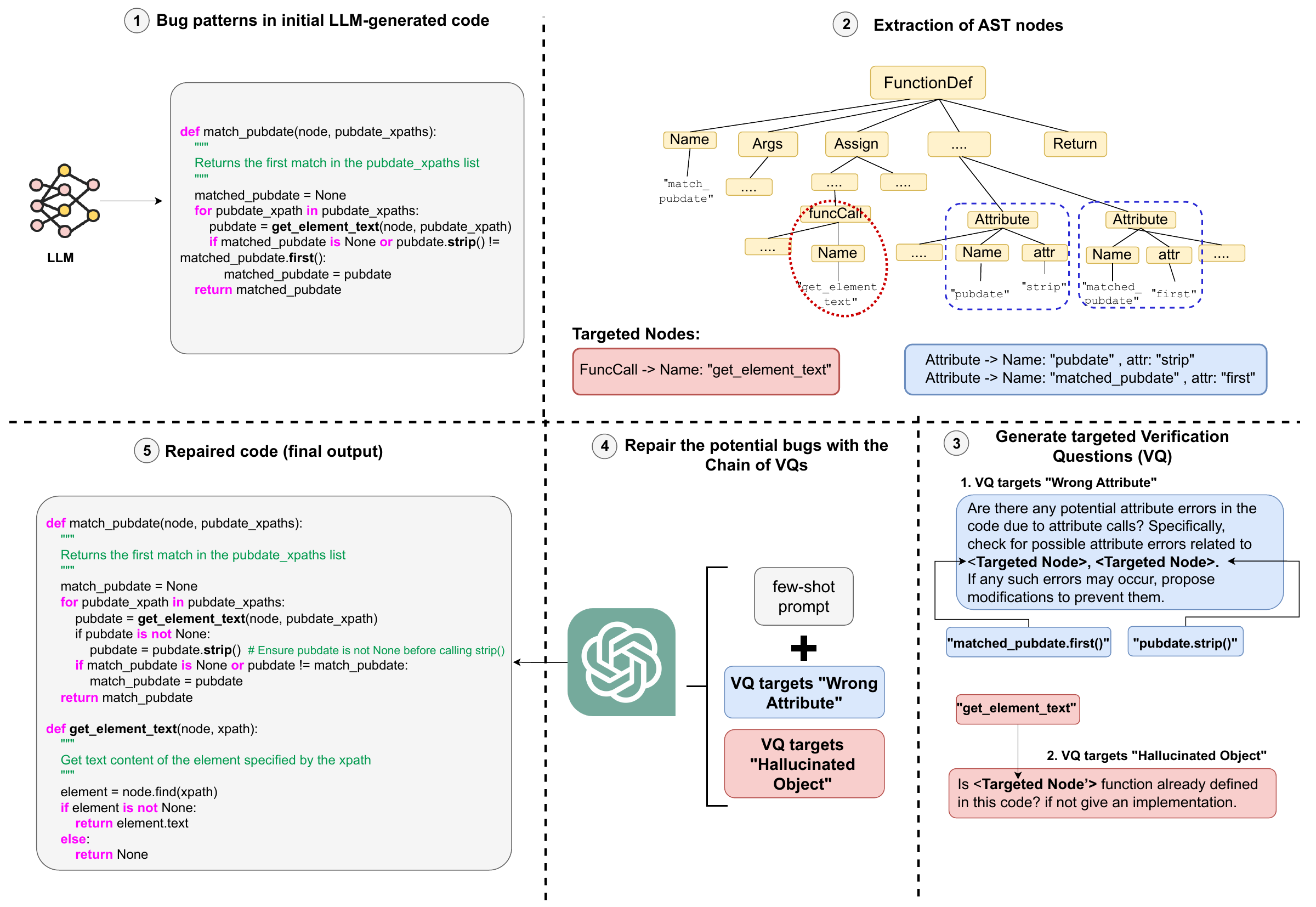}}
\caption{The proposed methodology for improving the reliability of code generated by LLMs in the absence of test cases.}
\label{fig-methodology}
\vspace{-1em}
\end{figure*}

Our methodology is divided into four steps: \textcircled{\raisebox{-.9pt}{1}} First, the user leverages LLM to generate a code based on the prompt of the task, \textcircled{\raisebox{-.9pt}{2}} the method parse the obtained code AST, collecting features from relevant nodes, \textcircled{\raisebox{-.9pt}{3}} based on the templates of VQs, targeted questions are generated by incorporating features from previously extracted nodes, \textcircled{\raisebox{-.9pt}{4}} finally, combining the questions with a few-shot prompt, the method queries ChatGPT on the specific task to obtain a repaired code which is presented to the user. A schematic description of the method is presented in Figure \ref{fig-methodology}. In the following, we describe the different steps in more detail.

\subsection{Initial prompt}
The method starts by prompting the LLM to generate code for a specific programming task and collect its initial output. The prompt consists of a description of the programming task in natural language. 
Figure~\ref{fig-methodology} \textcircled{\raisebox{-.9pt}{1}} illustrates an example of a function named ``\textit{match\_pubdate}'' generated by 
CodeGen in Python for a programming task described as ``\textit{Returns the first match in the pubdate\_xpaths list}''. This function aims to return the first node of an XML object that has a match in a given list of nodes. While our methodology is applicable across different programming languages, this study primarily focuses on the code generated by LLMs in Python as it is a very common and widely used programming language across different domains~\cite{python23}. 

\subsection{Extraction of AST nodes}

The Abstract Syntax Tree (AST) is an abstract representation of the parsing tree of code written in a programming language~\cite{cui2010code}. It is a tree data structure that captures the syntactic structure of the code while abstracting away details like punctuation and semicolons. Each node in the tree represents a syntactic construct of the code, such as expressions, statements, or declarations, and the tree's structure reflects the hierarchical relationships between these constructs. For example, Figure~\ref{fig-methodology} \textcircled{\raisebox{-.9pt}{2}} shows a part of the AST dump of the function ``\textit{match\_pubdate}'' in Figure~\ref{fig-methodology} \textcircled{\raisebox{-.9pt}{1}}. Each node in the AST also contains various features. For instance, the ``\textit{Attribute}'' node in the AST shown in Figure~\ref{fig-methodology} \textcircled{\raisebox{-.9pt}{2}} includes different features such as \textit{name} and \textit{attr}.

In the second step, the method localizes potential bugs by collecting different features from the initial LLM-generated code that may trigger specific types of errors. To localize the potential bugs, the method walks through the AST of the initial LLM-generated code and collects features on some targeted nodes that may trigger specific types of errors.

\subsubsection{\textbf{Bug patterns in LLM-generated code}}
 Within the various types of bugs observed in LLM-generated code in Python, in this study, we focus on two specific categories identified in the LLM-generated codes' bugs taxonomy proposed by Tambon et al. \cite{tambon2024bugs}: ``\textit{Wrong Attribut}'' and ``\textit{Hallucinated Object}''.  We narrowed down our investigation to these two types because, as indicated by the survey study on this taxonomy, participants highlighted them as patterns that are easy to detect but challenging to repair. However, our proposed method can be extended to address other bug patterns as well.

 In the code example in Figure~\ref{fig-methodology} \textcircled{\raisebox{-.9pt}{1}}, in the \textit{if} condition, there are two attribute calls, ``\textit{strip()}'' and ``\textit{first()}'' (highlighted in bold). Such attribute calls hold the potential to trigger \textbf{\textit{Attribute Error}} if they are invalid. In the absence of test cases, identifying whether these attribute calls can be accurately applied or if they might raise errors becomes challenging. For instance, in Figure~\ref{fig-methodology} \textcircled{\raisebox{-.9pt}{1}}, ``\textit{strip()}'' correctly operates as an attribute call on the variable ``\textit{pubdate}''. However, ``\textit{first()}'' is not a valid attribute in Python to be applied to the variable ``\textit{matched\_pubdate}''. According to the taxonomy of bug patterns in LLM-generated code~\cite{tambon2024bugs}, this type of bug is considered as ``\textit{Wrong Attribute}''.

The code obtained in Figure~\ref{fig-methodology} \textcircled{\raisebox{-.9pt}{1}} also incorporates a function called ``\textit{get\_element\_text}'' (highlighted in bold). Such function calls hold the potential to trigger \textbf{\textit{Name Error}} if they are not defined earlier. In this example, the model generates the code with the assumption that the ``\textit{get\_element\_text}'' function should either be pre-defined or could be implemented. This kind of bug, while easy to notice, is hard to fix. In fact, this type of bug essentially requires implementing the primary objective of the task, which the model assumes is fulfilled by the hallucinated function. According to the taxonomy of bug patterns in LLM-generated code~\cite{tambon2024bugs}, this type of bug is classified as a ``\textit{Hallucinated Object}''.  

These two bug patterns can be linked to specific nodes in the AST. For example, the bug pattern ``\textit{Wrong Attribute}'' which can be identified by an \textit{Attribute Error} in Python, is linked to the node ``\textit{Attribute}'' in AST and the bug patterns ``\textit{Hallucinated Object}'' which can raise \textit{Name Error} in Python, is linked to the node ``\textit{Call}'' with \textit{func} feature. The colored box in Figure~\ref{fig-methodology} \textcircled{\raisebox{-.9pt}{2}} shows the features collected by our method on these targeted nodes.



\subsection{Generating targeted Verification Questions (VQs)}\label{sec:VQ}
In this step, the method generates targeted Verification Questions (VQs) by considering features collected from each targeted node— nodes with the potential to trigger specific types of error. To do so, we first design templates for VQs in a way to verify the correctness of features of the targeted nodes or repair them if they are inaccurate.

To obtain an appropriate VQ template for each targeted node,  we began by manually crafting VQs based on the bug patterns in LLM-generated code and the relevant nodes extracted from its AST. For instance, we craft a template designed to check the ``\textit{Wrong Attribute}'' pattern and verify attribute calls for potential \textit{Attribute Error}s. The template VQ is ``\textit{Can you verify that the following attribute calls will not generate attribute error: <Targeted Nodes>. If any attribute error may occur, repair the code.}'', where \textit{<Targeted Nodes>} will be replaced by a list of relevant features extracted from the ``\textit{Attribute}'' nodes in the AST of the generated code.

To mitigate the bias in our hand-crafted VQs, we then leveraged ChatGPT (gpt-3.5-turbo)~\cite{chatgpt22} to generate final template VQs. To do so, we considered the hand-crafted VQs in a few-shot prompt and asked ChatGPT to generate similar template VQs. The few-shot prompt that we used in this step of the method can be found in our replication package~\cite{replication24}. This gives us general templates for the specific bug patterns we aim to repair. Figure~\ref{fig-methodology} \textcircled{\raisebox{-.9pt}{3}} shows the template VQs for two bug patterns in our study: ``\textit{Wrong Attribute}'' and ``\textit{Hallucniated Object}'', generated by ChatGPT. However, the targeted VQs can easily be expanded to other bug patterns in LLM-generated code by incorporating new VQ templates relevant to other bug patterns in LLM-generated code. 

Our method used the template VQs to generate the targeted VQs by replacing the \textit{<Targeted Nodes>} with the list of features collected from the AST of the LLM-generated Code, as shown in Figure \ref{fig-methodology} \textcircled{\raisebox{-.9pt}{3}}. For instance, in the VQ that targets ``\textit{Wrong Attribute}'', the \textit{<Targeted Nodes>} will be replaced by node features \textit{matched\_pubdate.first()} and \textit{pubdate.strip()} to generate the targeted VQ.  For the VQ that targets ``\textit{Hallucinated Object}'', the \textit{<Targeted Node>} will be replaced by \textit{get\_element\_text()} which is an undefined function in this example and represents a hallucination generated by the LLM.

\subsubsection{\textbf{Rephrasing templates of targeted VQs}}\label{sec:rephrase}
\begin{figure}[htbp]
\caption{Instruction prompt to rephrase template VQ for ``\textit{Wrong Attribute}'' bug pattern.}
\centering
\includegraphics[width=0.8\linewidth]{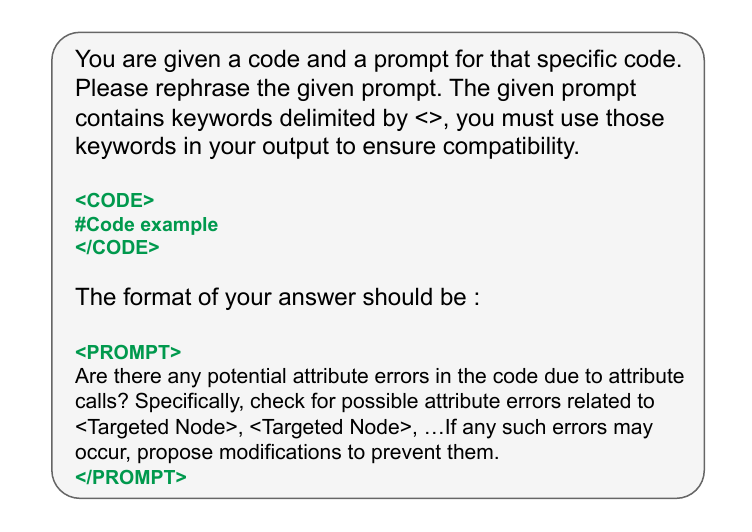}
\label{fig:rephAttr}
\end{figure}

Studies indicate that rephrasing a prompt into semantically equivalent prompts can impact the output generated by LLMs~\cite{mastropaolo2023robustness,rephrasing_paper}. Even slight modifications, such as altering words with seemingly synonymous meanings, can lead to variations in LM outputs. To investigate the role of template VQs in our proposed method, we rephrase them by leveraging ChatGPT~\cite{chatgpt22} for this rephrasing task. Drawing from previous research \cite{rephrasing_paper}, we employed the ChatGPT and experimented with different instructions to prompt the model for the rephrasing task. For instance, Figure~\ref{fig:rephAttr} represents the instruction that we used, inspired by Mizrahi et al. \cite{rephrasing_paper}, to rephrase the template VQ relevant to ``\textit{Wrong Attribute}'' bug pattern. Other instructions that we used in rephrasing the template VQs, along with the resulting rephrased template VQs are presented in our replication package~\cite{replication24}.

\vspace{30pt}

\subsection{Repair the potential bugs with the chain of VQs on LLM}\label{sec:repair}
In the final step, the method aggregates all the targeted VQs generated in Section~\ref{sec:VQ} into a chain of questions and attempts to use ChatGPT (gpt-3.5-turbo) to repair any bugs present in the initial LLM-generated code. 
We choose to use ChatGPT for repairing (so a different LLM than those that generated the initial code) as it allows for comparison across codes initially generated by different LLMs. Moreover, as the focus is on repairing an already generated code via targeted questions, leveraging a general model such as ChatGPT should yield better results than code-oriented LLMs. 

To prevent unnecessary alterations by the model when addressing the VQs or to avoid changing the correct targeted node, we employed a few-shot prompting strategy in this step (Figure \ref{fig-methodology} \textcircled{\raisebox{-.9pt}{4}}). The few-shot prompts 
include two samples of buggy code fragments (tagged by \textit{<CODE>}), their relevant VQs (tagged by \textit{<QUESTION>}), and the fixed code repaired only on the targeted nodes mentioned in the VQs (tagged by \textit{<CORRECTION>}). These examples guide the model to focus on repairing any bugs present in the targeted nodes while leaving them unchanged if they are already correct. The last code fragment in the few-shot prompt 
represents the code under verification. While the examples in the few-shot prompt remain constant, the last code fragment and its relevant VQs will be updated with the new buggy code generated by the LLM and its corresponding VQs generated in step 3, Section~\ref{sec:VQ}. The few-shot prompt, along with its corresponding fixed examples used in our method, is already included in the replication package~\cite{replication24}.

After invoking the prompt on ChatGPT, we expect the model to follow the VQs and provide, as a final output, a corrected version of the buggy code fragments. Figure \ref{fig-methodology} \textcircled{\raisebox{-.9pt}{5}} illustrates a repaired version of the buggy code in Figure \ref{fig-methodology} \textcircled{\raisebox{-.9pt}{1}}, generated by ChatGPT using the chain of VQs prompt. It is noteworthy that while the final output may not be entirely correct, it has been already repaired on the targeted nodes that are buggy. For example, in Figure \ref{fig-methodology} \textcircled{\raisebox{-.9pt}{5}},  \textit{pubdate.strip()} (considered in the VQ targeting the ``\textit{Wrong Attribute}'') remains unchanged as it was correct while \textit{matched\_pubdate.first()} has been corrected since it was identified as buggy. Also, as shown in Figure \ref{fig-methodology} \textcircled{\raisebox{-.9pt}{5}}, the model has generated a function \textit{get\_element\_text()} to repair the ``\textit{hallucinated object}'' problem presented in Figure \ref{fig-methodology} \textcircled{\raisebox{-.9pt}{1}}.

\section{Experiments}\label{sec:exp}

\subsection{Dataset and environment setup}\label{sec:data_and_environment}

We used the CoderEval~\cite{yu2023codereval} dataset in our experiments. The dataset is composed of 230 Python and 230 Java functions that were extracted from existing projects on GitHub. For each of those functions, the dataset contains the function's signature, the associated docstring, the function context (e.g., the complete file from which the function was extracted), and the actual extracted function which serves as an oracle. The dataset also contains 10 code samples, for each function, generated by three different LLMs using the previously mentioned information. In our experiments, we focused on the Python tasks which had some of the generated samples labeled as ``\textit{Wrong Attribute}'' or ``\textit{Hallucinated Object}'' in the taxonomy by Tambon et al. \cite{tambon2024bugs} as those are the bug patterns we aim to correct. This gives us 36 individual tasks.

In CoderEval, to properly assess the validity of a generated code sample, a Docker environment was set which contains all complete project functions used in the dataset. To assess the validity of a generated code sample, the user just has to insert the generated code sample in the corresponding project within the environment and run the available tests. On average, a task in CoderEval has 4.5 tests with a minimum of 1 and a maximum of 27. Running the tests allowed for measuring whether the generated code was correct or not. For the \textbf{36} tasks we gathered, we recovered \textbf{61} buggy samples affected by the ``\textit{Wrong Attribute}'' or ``\textit{Hallucinated Object}'' bug patterns and \textbf{54} correct samples as flagged in the CoderEval replication package thanks to those tests. Those samples will be used as the initial code fragments in our experiments.

\subsection{Experimental Setup}

To answer each of our research questions, we design the following experiments:

\subsubsection{RQ1: \textbf{\rqone}}

In this research question, we aim to assess whether the targeted VQs reduce the errors in the generated code. We will compare the results obtained with our method to two other methods: 1) a baseline without any verification questions, which simply lets the LLM generate the code without any refinement step (referred to as ``No VQ'' in the following), 2) another method has a refinement step based on asking a general verification question such as ``\textit{Can you improve this code or correct its bugs please ?}'' inspired by existing approaches \cite{fan2023automated, xia2023automated} (referred to as ``General VQ'' in the following).

To compare the methods, we apply each of the methods to the 61 buggy samples of each task of CoderEval as we described in Section \ref{sec:data_and_environment}. We repeat the process five times each time with a different seed to obtain a fair comparison between methods and account for the stochasticity of the generation process. We then collected the generated code fragments and inserted them in the CoderEval environment as described in Section \ref{sec:data_and_environment}. Finally, we ran all the available tests. We collected whether or not each individual test ran on the generated code (whether or not there is an ``AssertionError'') or if an exception was raised (e.g., ``AttributeError''). Thus, for each sample and each method, we have the results of each individual test which we split into: ``Runnable'' (the code runs whether or not there is an ``AssertionError''. This indicates that the code either successfully passed the test or it executed but failed to produce the expected output), ``Attribute errors'', ``Name errors'' and ``Other errors'' depending on the test outcome. It's worth noting that the ``Other errors'' category does not cover ``Assertion errors'', ``Attribute errors'' and ``Name errors''. As different tasks have different numbers of tests, we normalize the results per the number of tests. Furthermore, we aggregate those results at the sample level by considering the test results for a sample as follows: when the sample code runs for all tests (whether or not there are ``AssertionErrors''), we consider the sample as ``Runnable'' and if at least one of the test yields an exception, we consider the sample of the relevant error type category. Note that, as we are automatically executing tests, samples manually flagged as ``\textit{Hallucinated Object}'' and ``\textit{Wrong Attribute}'' in the taxonomy, would only count towards one error type, as the test execution would raise and stop at one error (i.e. the earliest one) before executing the rest of the code sample and raising other errors in the code sample. In that case, we consider only the first raise error for the error type of that code sample in our experiments.

\subsubsection{\textbf{RQ2: \rqtwo}}

The goal of this question is to study the impact of our methods on correct code in more detail. That is, to see if the method does not introduce defects in the correct code. To answer this question, we took from the CoderEval dataset the 54 correct code samples generated by LLMs for the  36 individual tasks considered. We applied our described methodology to those code fragments and ran the tests on the obtained generated code after applying our methodology. Similarly to RQ1, we executed five runs by varying the seed and averaging the obtained results. In each case, we checked whether a correct code ended up being changed wrongly, that is, the method introduced a bug in the newly generated code.

\subsubsection{\textbf{RQ3: \rqthree}}

The goal of this question is to study the impact of rephrasing the template of VQs on the output. We aim to assess if the repairing capability of the method is not hampered by how the templates of VQs are phrased. To do so, we generated five rephrasing versions of our initial template questions as explained in subsection~\ref{sec:rephrase} to see how rephrasing may affect the result. 
We used the same experiment process as RQ1 with our method by applying each of the rephrased prompts on the buggy samples.

\subsection{Results}

In the following, we present the results to answer each of our research questions.

\subsubsection {RQ1: \rqone}
Results at the test-level are presented in Table \ref{tab:test-level} and results for the sample level are given in Table \ref{tab:task-level}. ``No VQ'' and ``General VQ'' are the baseline methods and ``Targeted VQs'' presents the results of our proposed method.

\begin{table} [h]
      \caption{Average number of test cases resulting in a given error category. Results are normalized by the number of tests for a single task.}
      \vspace{-10pt}
      \label{tab:test-level}
      \begin{tabular}{cccc}
        \toprule
        criteria & No VQ &  General VQ & Targeted VQs\\
        \midrule
        Runnable cases & 0.2 &  22.42 & \textbf{25.42}  \\
        Attribute errors& 17.13 & 9.74 & \textbf{5.8} \\
        Name errors & 18.13&  6.24 & \textbf{2.57} \\
        Other errors & 25.54& \textbf{22.6} &27.21 \\
      \bottomrule
    \end{tabular}
\end{table}

\begin{table}[h]
      \caption{Average number of samples resulting in a given error category}
      \vspace{-10pt}
      \label{tab:task-level}
      \begin{tabular}{cccc}
        \toprule
        criteria & No VQ &  General VQ & Targeted VQs\\
        \midrule
        Runnable code & 0 &  22.32 
 & \textbf{25.33}  \\
        Attribute errors& 16 &  9.64 & \textbf{5.7} \\
        Name errors & 17 &  6.24 &\textbf{2.39}  \\
        Other errors & 28 & \textbf{22.8} & 27.58 \\
      \bottomrule
    \end{tabular}
\end{table}

As observed in Table \ref{tab:test-level} and Table \ref{tab:task-level}, our method, ``Targeted VQs'' performs better repair compared to the baseline, ``No VQ'' (before refinement): the number of tests not leading to an error type drastically diminishes for the targeted error types, "Attribute errors" and "Name errors", to $5.8$ and $2.57$ respectively. It also slightly diminishes the number of "Other errors". Moreover, the number of runnable tests increases from $0.2$ for the ``No VQ'' method to $25.42$ for the ``Targeted VQs'' method. We obtain similar trends at the sample level. Note that, at the test-level (Table \ref{tab:test-level}), we obtain some tests that run while the sample code is buggy. This is because, for a few samples, the test case would trigger a branch that would not contain the bug. However, the sample itself remains buggy.

Our method, ``Targeted VQ'' shows improvement compared to the ``General VQ'' as well, mainly in terms of the targeted errors: "Attribute error" and "Name error". For example at the test level, the average number of tests with "Attribute errors" decreases from $9.74$ to $5.8$, and it decreases from $6.24$ to $2.57$ for the "Name errors". The number of "Other errors" increases, for example from $22.8$ to $27.58$ at the sample level. However, this cannot be controlled by the targeted VQs and so is possible that this is a side effect of the targeted VQs that may induce new bugs while repairing targeted bugs. We study this side effect in more detail in RQ2. For the ``Runnable Code'' category for example at the sample level (Table 2), the average number of runnable samples increased from $22.32$ to $25.33$. In conclusion, our proposed method, ``Targeted VQ'', translates by less code prone to "Attribute errors" and "Name errors" and slightly more "Runnable" code which was the goal of the chain of targeted VQs.

\subsubsection{RQ2: \rqtwo} As shown in the results in Table~\ref{tab:task-level-correct}, on average $6$ correct code samples out of the $54$ correct samples in our dataset turned to buggy ones by applying the chain of targeted VQs, which is a 12\% false positive rate. Those errors are mainly either ``other errors'' (other than the targeted errors), 3.8 on average, or ``Assertion errors'' (the code samples execute correctly but fail on at least one of the tests), 2 on average. The number of code that ends up having an "Attribute" or "Name" error is very low, 0.2 on average. As such, the chain of targeted VQs with a few-shot prompt as explained in subsection~\ref{sec:repair}, does not introduce many errors and, in particular, does not introduce the targeted errors. This demonstrates that while our method diminishes the number of targeted errors as indicated by the results in RQ1, in the absence of test cases where we are unable to discern the correctness of LLM-generated code, it will not erroneously turn too many correct code samples to the buggy ones (resulting in low false positives).

\begin{table}[h]
\vspace{-10pt}
      \caption{Average number of correct codes which are wrongly changed by our method. Results are given at sample level.}
      \label{tab:task-level-correct}
      \begin{tabular}{cc}
        \toprule
        Error Type & Average \# of samples\\
        \midrule
        From correct code to Assertion errors & 2 \\
        From correct code to Attribute errors& 0.2\\
        From correct code to Name errors & 0.2 \\
        From correct code to Other errors & 3.8\\
        Staying correct & 47.8  \\
      \bottomrule
    \end{tabular}
    \vspace{-10pt}
\end{table}

\subsubsection{RQ3: \rqthree}

Table \ref{tab:rephrasing-results} shows the results of the impact of rephrasing at the sample level. We can observe that the different rephrasing introduces variability in the results. The sensibility of the LLMs to the prompt was already observed in previous studies ~\cite{mastropaolo2023robustness,rephrasing_paper}. However, such rephrasing does not introduce a high variability in the results: for instance, the average number of runnable samples remains 25 while the average number of samples with "Attribute" or "Name" errors remains around 6 and 2.6 respectively. As such, while not immune to prompt rephrasing, our method is not highly impacted by the phrases in the prompt. 

\begin{table*}
\caption{Impact of rephrasing on the performances at a test normalized level evaluation}
\label{tab:rephrasing-results}
\begin{tabular}{cclll}
\toprule
criteria & Runnable cases & Attribute errors & Name errors & Other errors \\
\midrule
Rephrasing 0 & 25.42 & 5.8 &  2.57  & 27.21 \\
Rephrasing 1 & 25.04  & 5.31 &  2.84  & 27.81 \\
Rephrasing 2  & 24.85 & 6.66 &  2.42  & 27.07 \\
Rephrasing 3  & 24.99 & 5.73 &  2.69  & 27.59 \\
Rephrasing 4 & 25.09  & 6.47 &  2.83  & 26.31 \\
Rephrasing 5 & 25.79  & 7.09 &  2.3  & 25.82 \\

\bottomrule
\end{tabular}
\end{table*}

\section{Discussion}\label{sec:discuss}

Our study is the first study that focuses on bug patterns that are common in LLM-generated code, such as hallucination in function calls or attributes. It can localize potential bugs in LLM-generated code in the absence of test cases by compiling features from the AST of the generated code and then attempting to repair them using different instructions as a chain of targeted VQs. While bugs in LLM-generated code may result in the same error types (i.e. when using the Python interpreter) as those in human buggy code, the actions required to repair them differ from those applied to human buggy code. For example, a ``\textit{Hallucinated Object}'' is a type of bug pattern that triggers a \textit{Name Error} in Python. However, repairing this bug requires generating a new function that implements the primary objective of the task, similar to Figure~\ref{fig-methodology} (1), which is not a typical bug pattern seen in human-written code.

Our method is a self-refinement method that focuses primarily on enhancing the reliability of LLM-generated code by repairing its bugs before execution, without human intervention, and in the absence of test cases. This makes the approach particularly appealing compared to previous studies that are dependent on a set of comprehensive test cases or retrieval/fine-tuning-based methods~\cite{lewis2020retrieval,jin2023inferfix} which require a lot of data. On the contrary, our approach does not require any external data and relies on VQs which can be adapted for any kind of bug and the performance of LLMs in synthesizing code. Thus, our method can reduce the effort required for detecting and repairing bugs in LLM-generated code on the user's side in the absence of test cases and before execution. Moreover, it emphasizes trustworthiness towards the user, as it helps in preventing the said user from being confronted with buggy code~\cite{alshahwan2024assured}. 

Our proposed method, asking targeted VQs around potential bugs in code generated by LLM, is not limited to specific bug types studied in this paper and can be adapted to a diverse range of bug patterns. Our proposed method can be employed to generate a set of templates of targeted VQs for other bug patterns to enhance the reliability of LLM-generated code more comprehensively. Moreover, our method can be adapted for various improvements in LLM-generated code, including in quality, coding style, and comprehension. Our approach could particularly be helpful for those with less programming experience who may struggle to detect and repair bugs in code generated by the LLMs, sometimes overlooking certain errors~\cite{dakhel2023github}. 
Finally, the targeted VQs in our study focus on the bug itself without replacing the entire code. In our experiments, we noticed that the general VQ, one of the baselines, led the model to drastically alter the code instead of repairing a specific bug, sometimes replacing the buggy code with an entirely new code, while our proposed method enables more control over the changes induced in the code. 

However, our study only focused on the templates of VQs generated based on an initial predetermined structure and only investigated the effects of rephrasing on this template, as it was shown to have an impact on the LLMs output \cite{rephrasing_paper}. Further investigations are necessary to explore the implications of using different instructions within the template of VQs. Moreover. the templates are manually crafted before being adapted by ChatGPT. However, it would be possible to have an autonomous agent verifier comment to automatically generate and tailor the chain of VQs based on the initial generated code. Those considerations are left to future works.

\vspace{-10pt}
\section{Threats to Validity}\label{sec:ttv}

\textbf{\textit{Internal Validity:}} The format of the templates for VQs in our study is predetermined to target specific types of bugs and the internal validity is mainly due to whether the templates of targeted VQs had the intended effect. We employ ChatGPT to generate the final targeted VQs by using the predetermined structured template alongside a few-shot prompt to avoid potential bias of the VQ being crafted by a human. However, the predetermined structure and format of the VQ templates by a human might impose limitations on the VQs generated by ChatGPT. To mitigate the effect of VQs, we compared our approach to two baselines (no VQ and general VQ) and repeated the experiment with each sample five times to account for non-determinism in the generation. Moreover, as recent studies highlighted that prompt rephrasing could impact LLMs' results~\cite{mastropaolo2023robustness,rephrasing_paper}, we also studied the impact of rephrasing the templates of VQs on the results, but we didn't study the impact of different predetermined structured templates on the VQs generated by ChatGPT which can be discovered by future studies. The choice of the CoderEval dataset may also pose a threat to internal validity, given that it was chosen to leverage the taxonomy by Tambon et al. \cite{tambon2024bugs}, which serves as the foundation for the bug patterns examined in our study. However, the advantage of this benchmark lies in its real programming tasks and methods extracted from GitHub repositories. This aspect highlights the usefulness of our method on more practical programming tasks compared to those tasks in more classical benchmarks such as MBPP~\cite{austin2021program} and HumanEval~\cite{chen2021evaluating}.

\noindent{\textbf{\textit{Construct Validity:}}} As evaluation metrics, we consider the reduction of error types triggered by the targeted bug patterns in our study (\textit{Attribute Error} and \textit{Name Error}), along with the capacity of the code to be \textit{Runnable} – that is, the code compiled without errors when provided with proper test inputs, even if the output isn't entirely correct. Furthermore, we compared the results of our method with two other baselines using these evaluation metrics. Additionally, we investigated whether our method introduces bugs into correct code (i.e. a form of ``false positive'') by examining the errors present in originally correct code after the application of our method. However, further investigation within a more comprehensive study is necessary to compare the regression among different types of errors in different code fragments by applying our method and other baselines.


\noindent{\textbf{\textit{External Validity:}}} In our study, we mainly focus on code samples generated by three LLMs (PanGu-Coder, CodeGen, and Codex), as described in the CoderEval dataset. This could hamper the generalization of our study. However, we argue that our approach is model-independent, given its focus on the AST of the generated code. While the initial code fragments were generated by different LLMs in our study, our method relies only on ChatGPT for generating the final targeted VQs and repairing buggy code by addressing the chain of VQs. Our approach may benefit from ChatGPT's performance over other models used for the initial code generation, but future studies can explore alternative models to gauge their impact on our findings. 

In this study, our focus was on two bug patterns from Tambon et al. \cite{tambon2024bugs} taxonomy, namely ``\textit{Hallucinated Object}'' and ``\textit{Wrong Attribute}''. We chose these patterns because they rank among the most prevalent in the taxonomy. However, our proposed method can be extended to cover other bug types within its chain of VQs. Finally, we focused on code fragments generated in Python, given its widespread usage as one of the most prevalent programming languages, and these samples are used within the taxonomy. Nonetheless, we expect that similar targeted VQ could be effectively generated for other programming languages.


\noindent{\textbf{\textit{Reliability Validity:}}} In our study, we focused on code fragments from CoderEval that were labeled either as ``\textit{Hallucinated Object}'' or ``\textit{Wrong Attribute}'' in Tambon et al. \cite{tambon2024bugs}. We verified that the samples led to such bug patterns before using them based on the dataset available on the taxonomy\cite{tambon2024bugs}. To enable other researchers to replicate or expand upon our study, we provide a replication package~\cite{replication24}. However, the ongoing enhancement of LLMs might pose challenges to achieving an exact replication of our findings.

\vspace{-12pt}
\section{Related works}\label{sec:related}

Some of the previous studies used APR tools to repair the buggy code generated by LLMs in order to enhance the reliability of their generated code~\cite{{fan2023automated,dakhel2023github}}. For instance, Fan et al. ~\cite{fan2023automated} employed an APR tool to first detect and then repair the buggy code generated by Codex for specific LeetCode problems. Additionally, they re-prompt Codex with various single instruction prompts, such as ``provide a fix for the buggy program'' or ``fix line N'',  to repair the bug in its initial generated code. Their findings indicate that leveraging Codex to repair its buggy-generated code performs better than an APR tool. This finding could be attributed to the fact that current APR tools are tailored to human bug patterns and may not be entirely effective on LLM-generated buggy code. Other studies, subsequent to generating the initial code by LLMs, re-prompt the model by incorporating additional information about the bugs into the prompt, such as error messages ~\cite{zhang2209repairing,skreta2023errors,schafer2023adaptive,deligiannis2023fixing}. Zhang et al. ~\cite{Zhang2023critical} investigated the potential of ChatGPT for program repair. Their results, based on a dataset of programming tasks, demonstrate that integrating error type and error line into the prompt enhances the repair capability of ChatGPT compared to a general instruction prompt like ``repair the bug in the program''. Another study proposed a fully autonomous code generation method by leveraging LLMs and adding the failed test cases into the prompt~\cite{liventsev2023fully}. Their method first collects the test cases that failed on the code fragments generated by Codex. Then they construct a repair instruction by appending the input test to the prompt and re-prompting the LLM to repair the code in a manner that generates the expected test output based on the relevant test input. Their results outperform both the traditional genetic programming approaches and the initial output of Codex in generating correct code fragments. Another type of information that studies have shown to be useful for leveraging LLMs for program repairs is, adding relevant examples in the form of few-shot prompts~\cite{ahmed2023better,zhang2209repairing,jin2023inferfix}. For example,  Joshin et al. ~\cite{joshi2023repair} used the similarity between the error message of the target buggy program and the error message of a pool of buggy programs and their fixed versions, to collect relevant examples and added them into the prompt as a few-shot of buggy programs and their fixes. They employed a static analysis tool to localize the bug and compile the error messages. Their results demonstrate an improvement in repairing bugs in human code (not LLM-generated) compared to an APR tool. COCO~\cite{yan2023coco} investigated the robustness of LLMs by adding information collected from the AST of the code into the initial prompts. Due to a robust model, adding features collected from the AST of its initial generated code to the same prompt—such as adding the imported library used in the initially generated code—shouldn't alter the model's output and should maintain consistency with the semantics of the initial code. However, their results indicate that incorporating such information into the same prompt can indeed alter the output of the LLM and is a good technique to evaluate their robustness. A study by Dhuliawala et al. ~\cite{dhuliawala2023chain} aimed to reduce hallucination in the natural language content generated by LLMs by fact-checking various aspects of the generated responses. They accomplished this by posing verification questions about different facts in the generated responses, such as ``Where was Clinton born?'' Their method then independently posed these verification questions to the LLM and used the model's responses to adjust the initial response accordingly. Their method demonstrates an improvement in hallucinations in the responses generated by LLMs for certain questions.  Similarly, a study by Wu~\cite{wularge} also used LLMs to generate clarification questions for a programming task to improve the correctness of generated code. The method presented these clarification questions to the user and incorporated the user's responses to gain a better understanding of the programming task. In contrast, our method improves the reliability of LLM-generated code by repairing its bugs without human intervention and in the absence of test cases. While tools such as static analysis tools can detect a limited range of bug types, 
they may fail to identify the diverse range of error types in the absence of test cases, such as attribute errors in Python. 
Moreover, using both APR tools and error message information requires a comprehensive set of test cases to detect and repair the bug. 

\vspace{-10pt}
\section{Conclusion}\label{sec:concl}

In this study, we aim to enhance the reliability of code generated by different LLMs in the absence of test cases and without human intervention. We achieve this by generating a chain of targeted VQs and re-prompting an LLM to repair potential bugs present in the initially generated code. To localize the potential bugs in the code generated by an LLM, our method traverses the AST of the initial code and gathers features from nodes that may trigger specific bug patterns. Then, it incorporates those features into the templates of targeted VQs generated by LLM and re-prompts the model to repair potential bugs by addressing the chain of VQs. Our focus in this study is primarily on two bug patterns based on the taxonomy of bugs in LLM code: ``\textit{Wrong Attribute}'' and ``\textit{Hallucinated Object}''. However, our proposed method is not limited to these patterns and can be extended to address other bug patterns. Our findings demonstrate that employing the chain of targeted VQs to re-prompt the LLM can reduce the number of \textit{Attribute Error} and \textit{Name Error} in the initial LLM-generated code fragments by up to 40\% and 62\%, respectively, compared to alternative baselines. Furthermore, applying the chain of targeted VQs to code fragments that were originally correct resulted in less than a 12\% conversion to buggy code. Additionally, the rephrasing of targeted VQs did not significantly impact the results of our method.

In future works, we plan to extend the chain of targeted VQs to discover additional bug patterns in the code generated by LLMs. We also intend to investigate the effects of employing various structures for template VQs within our method. Furthermore, our proposed method can be integrated into future studies where an agent has been trained to enhance the reliability of LLM-generated code by generating and addressing targeted VQs autonomously in an adversarial setup.

\section*{Acknowledgement}
This work is partially supported by the Canadian Institute for Advanced Research (CIFAR), and the Natural Sciences and Engineering Research Council
of Canada (NSERC).

\bibliographystyle{ACM-Reference-Format}
\bibliography{sample-authordraft}


\begin{thebibliography}{46}


\ifx \showCODEN    \undefined \def \showCODEN     #1{\unskip}     \fi
\ifx \showDOI      \undefined \def \showDOI       #1{#1}\fi
\ifx \showISBNx    \undefined \def \showISBNx     #1{\unskip}     \fi
\ifx \showISBNxiii \undefined \def \showISBNxiii  #1{\unskip}     \fi
\ifx \showISSN     \undefined \def \showISSN      #1{\unskip}     \fi
\ifx \showLCCN     \undefined \def \showLCCN      #1{\unskip}     \fi
\ifx \shownote     \undefined \def \shownote      #1{#1}          \fi
\ifx \showarticletitle \undefined \def \showarticletitle #1{#1}   \fi
\ifx \showURL      \undefined \def \showURL       {\relax}        \fi
\providecommand\bibfield[2]{#2}
\providecommand\bibinfo[2]{#2}
\providecommand\natexlab[1]{#1}
\providecommand\showeprint[2][]{arXiv:#2}

\bibitem[pyt(2020)]%
        {python23}
 \bibinfo{year}{2020}\natexlab{}.
\newblock \bibinfo{title}{Top Programming Languages 2020}.
\newblock \bibinfo{howpublished}{\url{https://spectrum.ieee.org/top-programming-language-2020}}.
\newblock


\bibitem[rep(2024)]%
        {replication24}
 \bibinfo{year}{2024}\natexlab{}.
\newblock \bibinfo{title}{The Replication Package}.
\newblock \bibinfo{howpublished}{\url{https://github.com/ExpertiseLLM/Chain-Of-Targeted-AST-Verification-Questions}}.
\newblock


\bibitem[Achiam et~al\mbox{.}(2023)]%
        {achiam2023gpt}
\bibfield{author}{\bibinfo{person}{Josh Achiam}, \bibinfo{person}{Steven Adler}, \bibinfo{person}{Sandhini Agarwal}, \bibinfo{person}{Lama Ahmad}, \bibinfo{person}{Ilge Akkaya}, \bibinfo{person}{Florencia~Leoni Aleman}, \bibinfo{person}{Diogo Almeida}, \bibinfo{person}{Janko Altenschmidt}, \bibinfo{person}{Sam Altman}, \bibinfo{person}{Shyamal Anadkat}, {et~al\mbox{.}}} \bibinfo{year}{2023}\natexlab{}.
\newblock \showarticletitle{Gpt-4 technical report}.
\newblock \bibinfo{journal}{\emph{arXiv preprint arXiv:2303.08774}} (\bibinfo{year}{2023}).
\newblock


\bibitem[Agarwal et~al\mbox{.}(2024)]%
        {agarwal2024copilot}
\bibfield{author}{\bibinfo{person}{Anisha Agarwal}, \bibinfo{person}{Aaron Chan}, \bibinfo{person}{Shubham Chandel}, \bibinfo{person}{Jinu Jang}, \bibinfo{person}{Shaun Miller}, \bibinfo{person}{Roshanak~Zilouchian Moghaddam}, \bibinfo{person}{Yevhen Mohylevskyy}, \bibinfo{person}{Neel Sundaresan}, {and} \bibinfo{person}{Michele Tufano}.} \bibinfo{year}{2024}\natexlab{}.
\newblock \showarticletitle{Copilot Evaluation Harness: Evaluating LLM-Guided Software Programming}.
\newblock \bibinfo{journal}{\emph{arXiv preprint arXiv:2402.14261}} (\bibinfo{year}{2024}).
\newblock


\bibitem[Ahmed and Devanbu(2023)]%
        {ahmed2023better}
\bibfield{author}{\bibinfo{person}{Toufique Ahmed} {and} \bibinfo{person}{Premkumar Devanbu}.} \bibinfo{year}{2023}\natexlab{}.
\newblock \showarticletitle{Better patching using LLM prompting, via Self-Consistency}. In \bibinfo{booktitle}{\emph{2023 38th IEEE/ACM International Conference on Automated Software Engineering (ASE)}}. IEEE, \bibinfo{pages}{1742--1746}.
\newblock


\bibitem[Alshahwan et~al\mbox{.}(2024)]%
        {alshahwan2024assured}
\bibfield{author}{\bibinfo{person}{Nadia Alshahwan}, \bibinfo{person}{Mark Harman}, \bibinfo{person}{Inna Harper}, \bibinfo{person}{Alexandru Marginean}, \bibinfo{person}{Shubho Sengupta}, {and} \bibinfo{person}{Eddy Wang}.} \bibinfo{year}{2024}\natexlab{}.
\newblock \showarticletitle{Assured LLM-Based Software Engineering}.
\newblock \bibinfo{journal}{\emph{arXiv preprint arXiv:2402.04380}} (\bibinfo{year}{2024}).
\newblock


\bibitem[Austin et~al\mbox{.}(2021)]%
        {austin2021program}
\bibfield{author}{\bibinfo{person}{Jacob Austin}, \bibinfo{person}{Augustus Odena}, \bibinfo{person}{Maxwell Nye}, \bibinfo{person}{Maarten Bosma}, \bibinfo{person}{Henryk Michalewski}, \bibinfo{person}{David Dohan}, \bibinfo{person}{Ellen Jiang}, \bibinfo{person}{Carrie Cai}, \bibinfo{person}{Michael Terry}, \bibinfo{person}{Quoc Le}, {et~al\mbox{.}}} \bibinfo{year}{2021}\natexlab{}.
\newblock \showarticletitle{Program synthesis with large language models}.
\newblock \bibinfo{journal}{\emph{arXiv preprint arXiv:2108.07732}} (\bibinfo{year}{2021}).
\newblock


\bibitem[Chen et~al\mbox{.}(2021)]%
        {chen2021evaluating}
\bibfield{author}{\bibinfo{person}{Mark Chen}, \bibinfo{person}{Jerry Tworek}, \bibinfo{person}{Heewoo Jun}, \bibinfo{person}{Qiming Yuan}, \bibinfo{person}{Henrique Ponde de~Oliveira Pinto}, \bibinfo{person}{Jared Kaplan}, \bibinfo{person}{Harri Edwards}, \bibinfo{person}{Yuri Burda}, \bibinfo{person}{Nicholas Joseph}, \bibinfo{person}{Greg Brockman}, {et~al\mbox{.}}} \bibinfo{year}{2021}\natexlab{}.
\newblock \showarticletitle{Evaluating large language models trained on code}.
\newblock \bibinfo{journal}{\emph{arXiv preprint arXiv:2107.03374}} (\bibinfo{year}{2021}).
\newblock


\bibitem[Copilot-chat(2023)]%
        {Copilotchat}
Copilot-chat \bibinfo{year}{2023}\natexlab{}.
\newblock \bibinfo{title}{Copilot Chat}.
\newblock \bibinfo{howpublished}{\url{https://docs.github.com/en/copilot/github-copilot-chat}}.
\newblock


\bibitem[Cui et~al\mbox{.}(2010)]%
        {cui2010code}
\bibfield{author}{\bibinfo{person}{Baojiang Cui}, \bibinfo{person}{Jiansong Li}, \bibinfo{person}{Tao Guo}, \bibinfo{person}{Jianxin Wang}, {and} \bibinfo{person}{Ding Ma}.} \bibinfo{year}{2010}\natexlab{}.
\newblock \showarticletitle{Code comparison system based on abstract syntax tree}. In \bibinfo{booktitle}{\emph{2010 3rd IEEE International Conference on Broadband Network and Multimedia Technology (IC-BNMT)}}. IEEE, \bibinfo{pages}{668--673}.
\newblock


\bibitem[Dakhel et~al\mbox{.}(2023a)]%
        {dakhel2023github}
\bibfield{author}{\bibinfo{person}{Arghavan~Moradi Dakhel}, \bibinfo{person}{Vahid Majdinasab}, \bibinfo{person}{Amin Nikanjam}, \bibinfo{person}{Foutse Khomh}, \bibinfo{person}{Michel~C Desmarais}, {and} \bibinfo{person}{Zhen Ming~Jack Jiang}.} \bibinfo{year}{2023}\natexlab{a}.
\newblock \showarticletitle{Github copilot ai pair programmer: Asset or liability?}
\newblock \bibinfo{journal}{\emph{Journal of Systems and Software}}  \bibinfo{volume}{203} (\bibinfo{year}{2023}), \bibinfo{pages}{111734}.
\newblock


\bibitem[Dakhel et~al\mbox{.}(2023b)]%
        {dakhel2023effective}
\bibfield{author}{\bibinfo{person}{Arghavan~Moradi Dakhel}, \bibinfo{person}{Amin Nikanjam}, \bibinfo{person}{Vahid Majdinasab}, \bibinfo{person}{Foutse Khomh}, {and} \bibinfo{person}{Michel~C Desmarais}.} \bibinfo{year}{2023}\natexlab{b}.
\newblock \showarticletitle{Effective test generation using pre-trained large language models and mutation testing}.
\newblock \bibinfo{journal}{\emph{arXiv preprint arXiv:2308.16557}} (\bibinfo{year}{2023}).
\newblock


\bibitem[Deligiannis et~al\mbox{.}(2023)]%
        {deligiannis2023fixing}
\bibfield{author}{\bibinfo{person}{Pantazis Deligiannis}, \bibinfo{person}{Akash Lal}, \bibinfo{person}{Nikita Mehrotra}, {and} \bibinfo{person}{Aseem Rastogi}.} \bibinfo{year}{2023}\natexlab{}.
\newblock \showarticletitle{Fixing rust compilation errors using llms}.
\newblock \bibinfo{journal}{\emph{arXiv preprint arXiv:2308.05177}} (\bibinfo{year}{2023}).
\newblock


\bibitem[Dhuliawala et~al\mbox{.}(2023)]%
        {dhuliawala2023chain}
\bibfield{author}{\bibinfo{person}{Shehzaad Dhuliawala}, \bibinfo{person}{Mojtaba Komeili}, \bibinfo{person}{Jing Xu}, \bibinfo{person}{Roberta Raileanu}, \bibinfo{person}{Xian Li}, \bibinfo{person}{Asli Celikyilmaz}, {and} \bibinfo{person}{Jason Weston}.} \bibinfo{year}{2023}\natexlab{}.
\newblock \showarticletitle{Chain-of-verification reduces hallucination in large language models}.
\newblock \bibinfo{journal}{\emph{arXiv preprint arXiv:2309.11495}} (\bibinfo{year}{2023}).
\newblock


\bibitem[Ebert et~al\mbox{.}(2019)]%
        {ebert2019confusion}
\bibfield{author}{\bibinfo{person}{Felipe Ebert}, \bibinfo{person}{Fernando Castor}, \bibinfo{person}{Nicole Novielli}, {and} \bibinfo{person}{Alexander Serebrenik}.} \bibinfo{year}{2019}\natexlab{}.
\newblock \showarticletitle{Confusion in code reviews: Reasons, impacts, and coping strategies}. In \bibinfo{booktitle}{\emph{2019 IEEE 26th international conference on software analysis, evolution and reengineering (SANER)}}. IEEE, \bibinfo{pages}{49--60}.
\newblock


\bibitem[Fan et~al\mbox{.}(2023)]%
        {fan2023automated}
\bibfield{author}{\bibinfo{person}{Zhiyu Fan}, \bibinfo{person}{Xiang Gao}, \bibinfo{person}{Martin Mirchev}, \bibinfo{person}{Abhik Roychoudhury}, {and} \bibinfo{person}{Shin~Hwei Tan}.} \bibinfo{year}{2023}\natexlab{}.
\newblock \showarticletitle{Automated repair of programs from large language models}. In \bibinfo{booktitle}{\emph{2023 IEEE/ACM 45th International Conference on Software Engineering (ICSE)}}. IEEE, \bibinfo{pages}{1469--1481}.
\newblock


\bibitem[GitHub-Copilot(2022)]%
        {GitHubCopilot}
GitHub-Copilot \bibinfo{year}{2022}\natexlab{}.
\newblock \bibinfo{title}{GitHub Copilot}.
\newblock \bibinfo{howpublished}{\url{https://github.com/features/copilot}}.
\newblock


\bibitem[Han et~al\mbox{.}(2021)]%
        {han2021understanding}
\bibfield{author}{\bibinfo{person}{Xiaofeng Han}, \bibinfo{person}{Amjed Tahir}, \bibinfo{person}{Peng Liang}, \bibinfo{person}{Steve Counsell}, {and} \bibinfo{person}{Yajing Luo}.} \bibinfo{year}{2021}\natexlab{}.
\newblock \showarticletitle{Understanding code smell detection via code review: A study of the openstack community}. In \bibinfo{booktitle}{\emph{2021 IEEE/ACM 29th International Conference on Program Comprehension (ICPC)}}. IEEE, \bibinfo{pages}{323--334}.
\newblock


\bibitem[Hassan et~al\mbox{.}(2024)]%
        {hassan2024rethinking}
\bibfield{author}{\bibinfo{person}{Ahmed~E Hassan}, \bibinfo{person}{Dayi Lin}, \bibinfo{person}{Gopi~Krishnan Rajbahadur}, \bibinfo{person}{Keheliya Gallaba}, \bibinfo{person}{Filipe~R Cogo}, \bibinfo{person}{Boyuan Chen}, \bibinfo{person}{Haoxiang Zhang}, \bibinfo{person}{Kishanthan Thangarajah}, \bibinfo{person}{Gustavo~Ansaldi Oliva}, \bibinfo{person}{Jiahuei Lin}, {et~al\mbox{.}}} \bibinfo{year}{2024}\natexlab{}.
\newblock \showarticletitle{Rethinking Software Engineering in the Era of Foundation Models: A Curated Catalogue of Challenges in the Development of Trustworthy FMware}.
\newblock \bibinfo{journal}{\emph{arXiv preprint arXiv:2402.15943}} (\bibinfo{year}{2024}).
\newblock


\bibitem[Jiang et~al\mbox{.}(2023)]%
        {jiang2023selfevolve}
\bibfield{author}{\bibinfo{person}{Shuyang Jiang}, \bibinfo{person}{Yuhao Wang}, {and} \bibinfo{person}{Yu Wang}.} \bibinfo{year}{2023}\natexlab{}.
\newblock \showarticletitle{SelfEvolve: A Code Evolution Framework via Large Language Models}.
\newblock \bibinfo{journal}{\emph{arXiv preprint arXiv:2306.02907}} (\bibinfo{year}{2023}).
\newblock


\bibitem[Jin et~al\mbox{.}(2023)]%
        {jin2023inferfix}
\bibfield{author}{\bibinfo{person}{Matthew Jin}, \bibinfo{person}{Syed Shahriar}, \bibinfo{person}{Michele Tufano}, \bibinfo{person}{Xin Shi}, \bibinfo{person}{Shuai Lu}, \bibinfo{person}{Neel Sundaresan}, {and} \bibinfo{person}{Alexey Svyatkovskiy}.} \bibinfo{year}{2023}\natexlab{}.
\newblock \showarticletitle{Inferfix: End-to-end program repair with llms}. In \bibinfo{booktitle}{\emph{Proceedings of the 31st ACM Joint European Software Engineering Conference and Symposium on the Foundations of Software Engineering}}. \bibinfo{pages}{1646--1656}.
\newblock


\bibitem[Joshi et~al\mbox{.}(2023)]%
        {joshi2023repair}
\bibfield{author}{\bibinfo{person}{Harshit Joshi}, \bibinfo{person}{Jos{\'e}~Cambronero Sanchez}, \bibinfo{person}{Sumit Gulwani}, \bibinfo{person}{Vu Le}, \bibinfo{person}{Gust Verbruggen}, {and} \bibinfo{person}{Ivan Radi{\v{c}}ek}.} \bibinfo{year}{2023}\natexlab{}.
\newblock \showarticletitle{Repair is nearly generation: Multilingual program repair with llms}. In \bibinfo{booktitle}{\emph{Proceedings of the AAAI Conference on Artificial Intelligence}}, Vol.~\bibinfo{volume}{37}. \bibinfo{pages}{5131--5140}.
\newblock


\bibitem[Kazemitabaar et~al\mbox{.}(2023)]%
        {kazemitabaar2023novices}
\bibfield{author}{\bibinfo{person}{Majeed Kazemitabaar}, \bibinfo{person}{Xinying Hou}, \bibinfo{person}{Austin Henley}, \bibinfo{person}{Barbara~Jane Ericson}, \bibinfo{person}{David Weintrop}, {and} \bibinfo{person}{Tovi Grossman}.} \bibinfo{year}{2023}\natexlab{}.
\newblock \showarticletitle{How novices use LLM-based code generators to solve CS1 coding tasks in a self-paced learning environment}. In \bibinfo{booktitle}{\emph{Proceedings of the 23rd Koli Calling International Conference on Computing Education Research}}. \bibinfo{pages}{1--12}.
\newblock


\bibitem[Lewis et~al\mbox{.}(2020)]%
        {lewis2020retrieval}
\bibfield{author}{\bibinfo{person}{Patrick Lewis}, \bibinfo{person}{Ethan Perez}, \bibinfo{person}{Aleksandra Piktus}, \bibinfo{person}{Fabio Petroni}, \bibinfo{person}{Vladimir Karpukhin}, \bibinfo{person}{Naman Goyal}, \bibinfo{person}{Heinrich K{\"u}ttler}, \bibinfo{person}{Mike Lewis}, \bibinfo{person}{Wen-tau Yih}, \bibinfo{person}{Tim Rockt{\"a}schel}, {et~al\mbox{.}}} \bibinfo{year}{2020}\natexlab{}.
\newblock \showarticletitle{Retrieval-augmented generation for knowledge-intensive nlp tasks}.
\newblock \bibinfo{journal}{\emph{Advances in Neural Information Processing Systems}}  \bibinfo{volume}{33} (\bibinfo{year}{2020}), \bibinfo{pages}{9459--9474}.
\newblock


\bibitem[Liu et~al\mbox{.}(2024)]%
        {liu2024your}
\bibfield{author}{\bibinfo{person}{Jiawei Liu}, \bibinfo{person}{Chunqiu~Steven Xia}, \bibinfo{person}{Yuyao Wang}, {and} \bibinfo{person}{Lingming Zhang}.} \bibinfo{year}{2024}\natexlab{}.
\newblock \showarticletitle{Is your code generated by chatgpt really correct? rigorous evaluation of large language models for code generation}.
\newblock \bibinfo{journal}{\emph{Advances in Neural Information Processing Systems}}  \bibinfo{volume}{36} (\bibinfo{year}{2024}).
\newblock


\bibitem[Liu et~al\mbox{.}(2023)]%
        {liu2023refining}
\bibfield{author}{\bibinfo{person}{Yue Liu}, \bibinfo{person}{Thanh Le-Cong}, \bibinfo{person}{Ratnadira Widyasari}, \bibinfo{person}{Chakkrit Tantithamthavorn}, \bibinfo{person}{Li Li}, \bibinfo{person}{Xuan-Bach~D Le}, {and} \bibinfo{person}{David Lo}.} \bibinfo{year}{2023}\natexlab{}.
\newblock \showarticletitle{Refining ChatGPT-generated code: Characterizing and mitigating code quality issues}.
\newblock \bibinfo{journal}{\emph{ACM Transactions on Software Engineering and Methodology}} (\bibinfo{year}{2023}).
\newblock


\bibitem[Liventsev et~al\mbox{.}(2023)]%
        {liventsev2023fully}
\bibfield{author}{\bibinfo{person}{Vadim Liventsev}, \bibinfo{person}{Anastasiia Grishina}, \bibinfo{person}{Aki H{\"a}rm{\"a}}, {and} \bibinfo{person}{Leon Moonen}.} \bibinfo{year}{2023}\natexlab{}.
\newblock \showarticletitle{Fully Autonomous Programming with Large Language Models}.
\newblock \bibinfo{journal}{\emph{arXiv preprint arXiv:2304.10423}} (\bibinfo{year}{2023}).
\newblock


\bibitem[Mastropaolo et~al\mbox{.}(2023)]%
        {mastropaolo2023robustness}
\bibfield{author}{\bibinfo{person}{Antonio Mastropaolo}, \bibinfo{person}{Luca Pascarella}, \bibinfo{person}{Emanuela Guglielmi}, \bibinfo{person}{Matteo Ciniselli}, \bibinfo{person}{Simone Scalabrino}, \bibinfo{person}{Rocco Oliveto}, {and} \bibinfo{person}{Gabriele Bavota}.} \bibinfo{year}{2023}\natexlab{}.
\newblock \showarticletitle{On the robustness of code generation techniques: An empirical study on github copilot}.
\newblock \bibinfo{journal}{\emph{arXiv preprint arXiv:2302.00438}} (\bibinfo{year}{2023}).
\newblock


\bibitem[Mizrahi et~al\mbox{.}(2024)]%
        {rephrasing_paper}
\bibfield{author}{\bibinfo{person}{Moran Mizrahi}, \bibinfo{person}{Guy Kaplan}, \bibinfo{person}{Dan Malkin}, \bibinfo{person}{Rotem Dror}, \bibinfo{person}{Dafna Shahaf}, {and} \bibinfo{person}{Gabriel Stanovsky}.} \bibinfo{year}{2024}\natexlab{}.
\newblock \bibinfo{title}{State of What Art? A Call for Multi-Prompt LLM Evaluation}.
\newblock
\newblock
\showeprint[arxiv]{2401.00595}~[cs.CL]


\bibitem[Sch{\"a}fer et~al\mbox{.}(2023)]%
        {schafer2023adaptive}
\bibfield{author}{\bibinfo{person}{Max Sch{\"a}fer}, \bibinfo{person}{Sarah Nadi}, \bibinfo{person}{Aryaz Eghbali}, {and} \bibinfo{person}{Frank Tip}.} \bibinfo{year}{2023}\natexlab{}.
\newblock \showarticletitle{Adaptive test generation using a large language model}.
\newblock \bibinfo{journal}{\emph{arXiv preprint arXiv:2302.06527}} (\bibinfo{year}{2023}).
\newblock


\bibitem[Schulman et~al\mbox{.}(2022)]%
        {chatgpt22}
\bibfield{author}{\bibinfo{person}{John Schulman}, \bibinfo{person}{Barret Zoph}, \bibinfo{person}{Christina Kim}, \bibinfo{person}{Jacob Hilton}, \bibinfo{person}{Jacob Menick}, \bibinfo{person}{Jiayi Weng}, \bibinfo{person}{Juan Felipe Ceron~Uribe}, \bibinfo{person}{Liam Fedus}, \bibinfo{person}{Luke Metz}, \bibinfo{person}{Michael Pokorny}, {et~al\mbox{.}}} \bibinfo{year}{2022}\natexlab{}.
\newblock \showarticletitle{ChatGPT: Optimizing Language Models for Dialogue}.
\newblock \bibinfo{journal}{\emph{OpenAI blog (2022)}} (\bibinfo{year}{2022}).
\newblock


\bibitem[Shaughnessy et~al\mbox{.}(2021)]%
        {shaughnessy2021think}
\bibfield{author}{\bibinfo{person}{Meghan Shaughnessy}, \bibinfo{person}{Rosalie DeFino}, \bibinfo{person}{Erin Pfaff}, {and} \bibinfo{person}{Merrie Blunk}.} \bibinfo{year}{2021}\natexlab{}.
\newblock \showarticletitle{I think I made a mistake: How do prospective teachers elicit the thinking of a student who has made a mistake?}
\newblock \bibinfo{journal}{\emph{Journal of Mathematics Teacher Education}}  \bibinfo{volume}{24} (\bibinfo{year}{2021}), \bibinfo{pages}{335--359}.
\newblock


\bibitem[Skreta et~al\mbox{.}(2023)]%
        {skreta2023errors}
\bibfield{author}{\bibinfo{person}{Marta Skreta}, \bibinfo{person}{Naruki Yoshikawa}, \bibinfo{person}{Sebastian Arellano-Rubach}, \bibinfo{person}{Zhi Ji}, \bibinfo{person}{Lasse~Bj{\o}rn Kristensen}, \bibinfo{person}{Kourosh Darvish}, \bibinfo{person}{Al{\'a}n Aspuru-Guzik}, \bibinfo{person}{Florian Shkurti}, {and} \bibinfo{person}{Animesh Garg}.} \bibinfo{year}{2023}\natexlab{}.
\newblock \showarticletitle{Errors are Useful Prompts: Instruction Guided Task Programming with Verifier-Assisted Iterative Prompting}.
\newblock \bibinfo{journal}{\emph{arXiv preprint arXiv:2303.14100}} (\bibinfo{year}{2023}).
\newblock


\bibitem[Spiess et~al\mbox{.}(2024)]%
        {spiess2024quality}
\bibfield{author}{\bibinfo{person}{Claudio Spiess}, \bibinfo{person}{David Gros}, \bibinfo{person}{Kunal~Suresh Pai}, \bibinfo{person}{Michael Pradel}, \bibinfo{person}{Md~Rafiqul~Islam Rabin}, \bibinfo{person}{Susmit Jha}, \bibinfo{person}{Prem Devanbu}, {and} \bibinfo{person}{Toufique Ahmed}.} \bibinfo{year}{2024}\natexlab{}.
\newblock \showarticletitle{Quality and Trust in LLM-generated Code}.
\newblock \bibinfo{journal}{\emph{arXiv preprint arXiv:2402.02047}} (\bibinfo{year}{2024}).
\newblock


\bibitem[Tambon et~al\mbox{.}(2024)]%
        {tambon2024bugs}
\bibfield{author}{\bibinfo{person}{Florian Tambon}, \bibinfo{person}{Arghavan~Moradi Dakhel}, \bibinfo{person}{Amin Nikanjam}, \bibinfo{person}{Foutse Khomh}, \bibinfo{person}{Michel~C Desmarais}, {and} \bibinfo{person}{Giuliano Antoniol}.} \bibinfo{year}{2024}\natexlab{}.
\newblock \showarticletitle{Bugs in Large Language Models Generated Code}.
\newblock \bibinfo{journal}{\emph{arXiv preprint arXiv:2403.08937}} (\bibinfo{year}{2024}).
\newblock


\bibitem[Touvron et~al\mbox{.}(2023)]%
        {touvron2023llama}
\bibfield{author}{\bibinfo{person}{Hugo Touvron}, \bibinfo{person}{Louis Martin}, \bibinfo{person}{Kevin Stone}, \bibinfo{person}{Peter Albert}, \bibinfo{person}{Amjad Almahairi}, \bibinfo{person}{Yasmine Babaei}, \bibinfo{person}{Nikolay Bashlykov}, \bibinfo{person}{Soumya Batra}, \bibinfo{person}{Prajjwal Bhargava}, \bibinfo{person}{Shruti Bhosale}, {et~al\mbox{.}}} \bibinfo{year}{2023}\natexlab{}.
\newblock \showarticletitle{Llama 2: Open foundation and fine-tuned chat models}.
\newblock \bibinfo{journal}{\emph{arXiv preprint arXiv:2307.09288}} (\bibinfo{year}{2023}).
\newblock


\bibitem[Tufano et~al\mbox{.}(2024)]%
        {tufano2024unveiling}
\bibfield{author}{\bibinfo{person}{Rosalia Tufano}, \bibinfo{person}{Antonio Mastropaolo}, \bibinfo{person}{Federica Pepe}, \bibinfo{person}{Ozren Dabi{\'c}}, \bibinfo{person}{Massimiliano Di~Penta}, {and} \bibinfo{person}{Gabriele Bavota}.} \bibinfo{year}{2024}\natexlab{}.
\newblock \showarticletitle{Unveiling ChatGPT's Usage in Open Source Projects: A Mining-based Study}.
\newblock \bibinfo{journal}{\emph{arXiv preprint arXiv:2402.16480}} (\bibinfo{year}{2024}).
\newblock


\bibitem[Vaithilingam et~al\mbox{.}(2022)]%
        {vaithilingam2022expectation}
\bibfield{author}{\bibinfo{person}{Priyan Vaithilingam}, \bibinfo{person}{Tianyi Zhang}, {and} \bibinfo{person}{Elena~L Glassman}.} \bibinfo{year}{2022}\natexlab{}.
\newblock \showarticletitle{Expectation vs. experience: Evaluating the usability of code generation tools powered by large language models}. In \bibinfo{booktitle}{\emph{Chi conference on human factors in computing systems extended abstracts}}. \bibinfo{pages}{1--7}.
\newblock


\bibitem[Wu({[n.\,d.]})]%
        {wularge}
\bibfield{author}{\bibinfo{person}{Jie~JW Wu}.} \bibinfo{year}{[n.\,d.]}\natexlab{}.
\newblock \showarticletitle{Large Language Models Should Ask Clarifying Questions to Increase Confidence in Generated Code}.
\newblock  (\bibinfo{year}{[n.\,d.]}).
\newblock


\bibitem[Wu et~al\mbox{.}(2023a)]%
        {wu2023autogen}
\bibfield{author}{\bibinfo{person}{Qingyun Wu}, \bibinfo{person}{Gagan Bansal}, \bibinfo{person}{Jieyu Zhang}, \bibinfo{person}{Yiran Wu}, \bibinfo{person}{Shaokun Zhang}, \bibinfo{person}{Erkang Zhu}, \bibinfo{person}{Beibin Li}, \bibinfo{person}{Li Jiang}, \bibinfo{person}{Xiaoyun Zhang}, {and} \bibinfo{person}{Chi Wang}.} \bibinfo{year}{2023}\natexlab{a}.
\newblock \showarticletitle{Autogen: Enabling next-gen llm applications via multi-agent conversation framework}.
\newblock \bibinfo{journal}{\emph{arXiv preprint arXiv:2308.08155}} (\bibinfo{year}{2023}).
\newblock


\bibitem[Wu et~al\mbox{.}(2023b)]%
        {wu2023brief}
\bibfield{author}{\bibinfo{person}{Tianyu Wu}, \bibinfo{person}{Shizhu He}, \bibinfo{person}{Jingping Liu}, \bibinfo{person}{Siqi Sun}, \bibinfo{person}{Kang Liu}, \bibinfo{person}{Qing-Long Han}, {and} \bibinfo{person}{Yang Tang}.} \bibinfo{year}{2023}\natexlab{b}.
\newblock \showarticletitle{A brief overview of ChatGPT: The history, status quo and potential future development}.
\newblock \bibinfo{journal}{\emph{IEEE/CAA Journal of Automatica Sinica}} \bibinfo{volume}{10}, \bibinfo{number}{5} (\bibinfo{year}{2023}), \bibinfo{pages}{1122--1136}.
\newblock


\bibitem[Xia et~al\mbox{.}(2023)]%
        {xia2023automated}
\bibfield{author}{\bibinfo{person}{Chunqiu~Steven Xia}, \bibinfo{person}{Yuxiang Wei}, {and} \bibinfo{person}{Lingming Zhang}.} \bibinfo{year}{2023}\natexlab{}.
\newblock \showarticletitle{Automated program repair in the era of large pre-trained language models}. In \bibinfo{booktitle}{\emph{2023 IEEE/ACM 45th International Conference on Software Engineering (ICSE)}}. IEEE, \bibinfo{pages}{1482--1494}.
\newblock


\bibitem[Yan et~al\mbox{.}(2023)]%
        {yan2023coco}
\bibfield{author}{\bibinfo{person}{Ming Yan}, \bibinfo{person}{Junjie Chen}, \bibinfo{person}{Jie~M Zhang}, \bibinfo{person}{Xuejie Cao}, \bibinfo{person}{Chen Yang}, {and} \bibinfo{person}{Mark Harman}.} \bibinfo{year}{2023}\natexlab{}.
\newblock \showarticletitle{Coco: Testing code generation systems via concretized instructions}.
\newblock \bibinfo{journal}{\emph{arXiv preprint arXiv:2308.13319}} (\bibinfo{year}{2023}).
\newblock


\bibitem[Yu et~al\mbox{.}(2023)]%
        {yu2023codereval}
\bibfield{author}{\bibinfo{person}{Hao Yu}, \bibinfo{person}{Bo Shen}, \bibinfo{person}{Dezhi Ran}, \bibinfo{person}{Jiaxin Zhang}, \bibinfo{person}{Qi Zhang}, \bibinfo{person}{Yuchi Ma}, \bibinfo{person}{Guangtai Liang}, \bibinfo{person}{Ying Li}, \bibinfo{person}{Tao Xie}, {and} \bibinfo{person}{Qianxiang Wang}.} \bibinfo{year}{2023}\natexlab{}.
\newblock \showarticletitle{CoderEval: A Benchmark of Pragmatic Code Generation with Generative Pre-trained Models}.
\newblock \bibinfo{journal}{\emph{arXiv preprint arXiv:2302.00288v1}} (\bibinfo{year}{2023}).
\newblock


\bibitem[Zhang et~al\mbox{.}({[n.\,d.]})]%
        {zhang2209repairing}
\bibfield{author}{\bibinfo{person}{Jialu Zhang}, \bibinfo{person}{Jos{\'e} Cambronero}, \bibinfo{person}{Sumit Gulwani}, \bibinfo{person}{Vu Le}, \bibinfo{person}{Ruzica Piskac}, \bibinfo{person}{Gustavo Soares}, {and} \bibinfo{person}{Gust Verbruggen}.} \bibinfo{year}{[n.\,d.]}\natexlab{}.
\newblock \showarticletitle{Repairing bugs in python assignments using large language models (2022)}.
\newblock \bibinfo{journal}{\emph{URL: https://arxiv. org/abs/2209.14876, doi}}  \bibinfo{volume}{10} (\bibinfo{year}{[n.\,d.]}).
\newblock


\bibitem[Zhang et~al\mbox{.}(2023)]%
        {Zhang2023critical}
\bibfield{author}{\bibinfo{person}{Quanjun Zhang}, \bibinfo{person}{Tongke Zhang}, \bibinfo{person}{Juan Zhai}, \bibinfo{person}{Chunrong Fang}, \bibinfo{person}{Bowen Yu}, \bibinfo{person}{Weisong Sun}, {and} \bibinfo{person}{Zhenyu Chen}.} \bibinfo{year}{2023}\natexlab{}.
\newblock \showarticletitle{A critical review of large language model on software engineering: An example from chatgpt and automated program repair}.
\newblock \bibinfo{journal}{\emph{arXiv preprint arXiv:2310.08879}} (\bibinfo{year}{2023}).
\newblock


\end{thebibliography}

\end{document}